\documentstyle[12pt,aps]{revtex}
\begin{document}
\baselineskip 0.75cm
\begin{center}
\vspace{2.3cm}
{\large{\bf The fermion dynamical symmetry model for the even--even \\
            and even--odd nuclei in the Xe--Ba region}}             \\
\vspace{0.8cm}
        { Xing-Wang Pan~, Jia-Lun Ping\footnote{Permanent address:
Physics Department, Nanjing Normal University, Nanjing, 210097,
P.R. China}, ~Da Hsuan Feng }    \\
{\small{\em Department of Physics, Drexel University,
            Philadelphia, PA 19104, USA}}                          \\
\vspace{0.5cm}
                  { Cheng-Li Wu }           \\
{\small{\em Department of Physics, Chung Yuan Christian University,
             Chung-Li, Taiwan 32023, ROC }}                          \\
\vspace{0.5cm}
                   {Michael W. Guidry }                  \\
{\small{\em Department of Physics, University of Tennessee,
            Knoxville, TN 37996-1200, USA; }}\\
{\small{\em Physics Division, Oak Ridge National Laboratory,
            Oak Ridge, TN 37831, USA; }}   \\

\vspace{0.8cm}

\today
\end{center}
\vspace{0.5cm}
\begin{center}
{\large{\bf Abstract}}
\end{center}
\noindent
The even--even and even--odd nuclei $^{126}$Xe-$^{132}$Xe and
$^{131}$Ba-$^{137}$Ba are shown to have a well-realized
$SO_8 \supset  SO_6 \supset
SO_3$ fermion dynamical symmetry. Their low-lying energy levels can be
described by a unified analytical expression with two (three)
adjustable parameters for
even--odd (even--even) nuclei that is
derived from the fermion dynamical symmetry
model. Analytical expressions are given for wavefunctions and for $E2$
transition rates that agree well with data.  The distinction between the
FDSM and IBM $SO_6$ limits is discussed. The experimentally observed
suppression of the the energy levels with increasing $SO_5$ quantum number
$\tau$ can be explained as a perturbation of the pairing interaction on the
$SO_6$ symmetry, which leads to an $SO_5$ Pairing effect for $SO_6$ nuclei.

\newpage

\renewcommand{\thesection}{\Roman{section}}

\section{Introduction}
In the past few decades, extremely rich experimental data have accumulated
for low-lying nuclear spectroscopy. The observed levels are interwoven in a
rich and complicated manner, and understanding them is a challenging
problem. The low-energy spectroscopy of even--even, medium-heavy and
heavy nuclei
can now be explained rather well by the interacting boson model ({\sc IBM})
\cite{IBM}. The first attempt to describe on the same footing
the spectroscopy of both even--even
and even--odd nuclei in an analytical way is due to Iachello \cite{Ia.80},
through the
concept of supersymmetry.

In this paper, we shall use the terminology {\sc NSUSY} to denote nuclear
dynamical supersymmetry. NSUSY is an outgrowth  of the phenomenological {\sc
IBM} that treats fermions and bosons as basic building blocks and identifies
the even--even and even--odd collective nuclear states as multiplets of a
higher
symmetry described by a supergroup $U(6/4)$ or $U(6/20)$
\cite{Ia.80,Ba.81,Ling.84}. The decomposition of $U(6/4)$ ($U(6/20)$)
contains $U^{B}(6)\times
U^{F}(4)$ ($U^{B}(6)\times U^{F}(20)$) with $U^{B}(6)$ referring to the six
bosons and $U^{F}(4)$ ($U^{F}(20)$) to the odd fermion moving in a single-j
level of $j=\frac{3}{2}$ (multi-j levels with $j=\frac{1}{2}$, $j=\frac{3}{2}$,
$j=\frac{5}{2}$ and $j=\frac{7}{2}$ ). Later, Jolie {\it et al.}
\cite{Jolie.87}
proposed a new reduction scheme for the supergroup $U(6/20)$ and applied it to
the $A=130$ mass region. They considered the single-particle orbits
$\frac{1}{2}$, $\frac{3}{2}$, $\frac{5}{2}$ and $\frac{7}{2}$ resulting from
the coupling of a pseudo orbital $l =2$ part and a pseudo spin $s=\frac{3}{2}$.
Instead of the group chain $U(6/20) \supset U^{B}(6)\times U^{F}(4)$
\cite{Ling.84}, they suggested use of the group chain
\begin{eqnarray}
& U(6/20)\supset U^{B}(6)\times U^{F}(20) \supset O^{B}(6)\times U^{F}(4)
\times U^{F}(5)\supset Spin(6) \times U^{F}(5) &   \nonumber   \\
& \supset Spin(5)\times O^{F}(5)\supset   \overline{Spin}(5)\supset Spin(3)~. &
\label{eq11}
\end{eqnarray}

{\sc NSUSY} has had some successes (generally up to 15$\sim$30\% accuracy
for spectroscopic fitting for a few nuclei). But the basic building blocks of
nuclei are fermions, and the $s$ and $d$ bosons in the {\sc IBM} are supposedly
simulations of coherent nucleon pairs with angular momenta 0 and 2.
Therefore, it is a simplification of the
real situation to introduce
$U^{B}(6)$ and $U^{F}(4)$ (or $U^{F}(5)$,
and $U^{F}(20)$)
as a direct product
$U^{B}(6) \times U^{F}(4)$ (or $U^{B}(6) \times U^{F}(20)$, or
$U^{B}(6) \times U^{F}(4)\times U^{F}(5)$).

The Fermion Dynamical Symmetry Model (FDSM) \cite{FDSM1,FDSM2} is defined
by a fermionic Lie algebra. It has symmetry limits analogous to all the IBM
limits and takes Pauli principle into
account \cite{Gi.80,Chen.86}. Furthermore,  the states
for even and odd systems
in the FDSM belong to vector and spinor representations, respectively,
of $SO_8$ or $Sp_6$; thus, one can
describe even--even and even--odd nuclei in the FDSM without
additional degrees of freedom.

Two general regions are though to exhibit some level of supersymmetry in the
properties of low-lying nuclear states: the Pt region
\cite{Ling.84,pto6} and the $A =130$ (Xe--Ba) region \cite{Ca.85}.
In the Pt region,  the normal-parity valence protons and neutrons are
in the 6th and 7th shells respectively, and according to the FDSM
\cite{FDSM2} they have $SO_8^{\pi} \times Sp_6^{\nu} $ symmetry,
which does not permit an analytical solution for the proton--neutron
coupled system. However it could have effective $SO_6$-like symmetry
as we have shown in\cite{SO6}. On the other hand, nuclei in the Xe--Ba
region have both their neutrons and their protons in the 6th shell
with FDSM pseudo-orbital angular momentum $k=2$ and pseudo-spin
$i=\frac{3}{2}$ for the normal-parity levels.
Thus they are expected to possess $SO^{\pi}_8\times SO^{\nu}_8$
symmetry, which contains coupled $SO_8$ symmetry (the FDSM analog
of IBM-1), and have
analytic solutions for the $SO_5 \times SU_2$,$SO_6$
and $SO_7$ dynamical symmetries. In fact, there is now
empirical evidence \cite{Ca.85} that nuclei in the $A =130$
region are better
portrayed by an $SO_6$ limit than the Pt region. For these reasons,
we have chosen the Xe--Ba region to discuss the possibility of a simplified
and unified
description for even--even and even--odd nuclei by the {\sc FDSM} as an
alternative to the IBM and to NSUSY.

The organization of the paper is as follows: energy formulas for both
even--even and even--odd nuclei and
 a comparison with data
are given in
Sec.\ II, the respective wavefunctions are constructed in Sec.\ III, the
electromagnetic transitions are discussed in Sec.\ IV, and
conclusions are presented in Sec.\ V.

\section{The energy spectra}
In the simplest version of the {\sc FDSM}, the numbers of nucleons in the
normal and abnormal orbits are fixed for a given nucleus
and therefore the quasi-spin group
${\cal SU}_{2}$  for the abnormal levels plays no explicit
dynamical
role for low-lying states (it enters implicitly through the conservation of
particle number and through effective interaction parameters).
The wavefunctions for both even and odd nuclei are given by the
following group chain
\begin{eqnarray}
  &(~SO^{i}_{8}~\supset~SO^{i}_{6}~\supset~ SO^{i}_{5}~)
 \times SO^{k}_{5}~~
  \supset~ SO^{i+k}_{5}~ \supset SO^{k+i}_{3}~~~~~~~~~~~~~~~ & \\
                                                \label{eq21}
  &~~~~[l_{1}l_{2}l_{3}l_{4}] \hspace {0.4cm}
 [\sigma_{1} \sigma_{2} \sigma_{3}]\hspace{0.4cm}
  [\tau_{1}\tau_{2}] \hspace{0.8cm} [\tau 0]\hspace{1.2cm}
  [\omega_{1}\omega_{2}]\hspace{0.8cm}
  J~~~~~~~~~~~~~~~~~~~~~~~&  \nonumber
\end{eqnarray}
where $[l_{1}l_{2}l_{3}l_{4}]$, $[\sigma_{1}\sigma_{2}\sigma_{3}]$
and $[\tau_{1}\tau_{2}]$ are the Cartan--Weyl labels for
the groups $SO_8$, $SO_6$ and $SO_5$, respectively, $\tau =0 (1)$
for even (odd) nuclei, and $k$ and $i$ indicate pseduo-orbital and
pseduo-spin parts of the groups respectively. We note the resemblance
between Eq.\ (2) and the {\sc NSUSY} group chain Eq.\ (\ref{eq11}).
The {\sc FDSM} Hamiltonian is
\begin{eqnarray}
 &H_{FDSM} =&  \varepsilon_{1}n_{1} + G_{0}S^{\dag}S
  + G_{2}D^{\dag}\cdot D + \sum_{r=1}^{3} B_{r} P^{r}(i) \cdot
  P^{r}(i)
\nonumber\\
 && +\sum_{r=1,3} \left[ {\cal B}_{r} P^{r}(k) \cdot P^{r}(k)
  +2b_{r} P^{r}(i) \cdot P^{r}(k) \right] ,
\label{eq22}
\end{eqnarray}
where $\varepsilon_{1}$ is the energy for the normal parity orbits
(assumed degenerate) and $n_{1}$ is the number of
nucleons in the normal parity orbits,
\begin{eqnarray}
\hspace{-1 in}   S^{\dag} & = & A^{0\dag} , ~~~~~D^{\dag}_{\mu} =
                  A^{2\dag}_{\mu} ,            \nonumber        \\
\hspace{-1 in}   A^{r\dag}_{\mu} & = & \sqrt{\Omega_{1}/2}
     \left[ b^{\dagger}_{ki} b^{\dagger}_{ki} \right]
                 ^{0r}_{0\mu} ,~~~~~~r=0,\, 2 ,
\label{eq23b}
\end{eqnarray}
where $k=2$, $i=\frac{3}{2}$ and $\Omega_{1}\equiv\Omega_{ki}
=(2k+1)(2i+1)/2$ . Similarly,
\begin{eqnarray}
\hspace{-0.1 in} P^{r}_{\mu}(i)&=& \sqrt{\Omega_{1}/2} \left[ b^{\dag}_{ki}
   \tilde{b} ^{\phantom{\dagger}}_{ki} \right] ^{0r}_{0\mu}, ~~~~
   r=0,1,2,3 ,\\
\hspace{-0.1 in} \bar{P}^{ r}_{\mu}(k)
 &=& (-)^{\left[\frac{r}{2}\right] } \sqrt{8/5} P^{r}_{\mu}(k),~~
P^{r}_{\mu}(k)=\sqrt{\Omega_{1}/2}\left[ b^{\dagger}_{ki}
   \tilde{b} ^{\phantom{\dag}}_{ki} \right] ^{r0}_{\mu 0}, ~~~~ r=0,1,2,3,
\label{eq24b}
\end{eqnarray}
where $\left[ \frac{r}{2}\right]$ is the integer part of
$\frac{r}{2}$. The operators
$P^r_{\mu}(i)$  and  $\bar{P}^r_{\mu}(k)$ for $r=1$ and 3
form the Lie algebras  $SO^i_5$ and $SO^k_5$,  respectively.
The commutators among the $P^r_{\mu}(i)$ are given by Eq.\ (3.12)
of ref.\ \cite{FDSM1} for the i-active case and those for $\sqrt{\Omega_1 /2}
[b^{\dag}_{ki} \tilde b_{ki}]^{r0}_{\mu 0}$
can be obtained from Eq.\ (3.12)
of ref.\ \cite{FDSM1} for the k-active case with
$$
\sqrt{3} \left\{ \begin{array}{ccc} r&s&t \\ 1&1&1 \end{array} \right\}
\longrightarrow \sqrt{5} \left\{ \begin{array}{ccc} r&s&t \\ 2&2&2
\end{array} \right\}.
$$
In Eq.\ (\ref{eq24b}) we
have renormalized the multipole operators $\bar P^r_{\mu}(k)$ so
that they are isomorphic with $P^r_{\mu}(i)$. Furthermore, $P^1_{\mu}(i)$
and $\bar P^1_{\mu}(k)$ are related to the total pseudo-orbital
angular momentum and pseudo spin by
\begin{eqnarray}
P^1_{\mu}(i) = {1\over {\sqrt 5}} I_{\mu} ,
  ~~~~~~ \bar P^1_{\mu}(k) = {1\over {\sqrt 5}} L_{\mu}	.
\label{eq24c}
\end{eqnarray}

By using the Casimir operators of $SO_8$, $SO_6$ and $SO_5$,
the Hamiltonian of Eq.\ (3) can be rewritten as
\begin{eqnarray}
&H_{FDSM} =&H_0 + \epsilon_1n_1 + g_S S^{\dag}\cdot S
        + g_6C_{SO_6^i}  + g_5 C_{SO_5^{k+i}}
\nonumber\\
      &&+ g^i_5 C_{SO_5^i} + g^k_5 C_{SO_5^k}
        + g_I {\bf I}^2 + g_L {\bf L}^2 + g_J {\bf J}^2 ,
\label{eq25a}
\end{eqnarray}
where the total angular momentum is $\bf J = I + L$, and
\begin{equation}
\begin{array}{rcl}
H_0 &=& {1\over 4}(n_1)^2 + G_2 [C_{SO_8^i} - S_0(S_0 - 6)], ~~~~
S_0 = {1\over 2} (n_1 - \Omega_1), \\
g_S &=& (G_0 - G_2), ~~~~ g_6 = (B_2 - G_2), ~~~~ g_5 = b_3, \\
g^i_5 &=& g'_5 - g_5, ~~~~ g'_5 = B_3 - B_2, ~~~~ g^k_5 = {\cal B}_3 - b_3,\\
g_I &=& g'_I - g_J , ~~~~  g'_I = {1\over 5} (B_1 - B_3),\\
g_J &=& {1\over 5} (b_1 + b_3), ~~~~ g_L = {1\over 5}
({\cal B}_1 - {\cal B}_3) -
{1\over 8} g_J.
\end{array}
\label{eq25b}
\end{equation}
The eigenvalue of $C_{SO_8}$ is
\begin{eqnarray}
C_{SO_8} = \sum^4_{i=1} l_i ~ (l_i + 8 - 2i) .
\label{eq25c}
\end{eqnarray}
and the condition for the realization of the symmetry is $g_S = 0$,
implying that
\begin{eqnarray}
G_0 = G_2 .
\label{eq26a}
\end{eqnarray}
The low-lying states of even--even nuclei belong to the $SO_8$ irrep
$[{\Omega_1\over 4} {\Omega_1\over 4} {\Omega_1\over 4} {\Omega_1\over 4}]$,
i.e., the irrep with
heritage $u = 0$, $u$ being the number of valence
nucleons not contained in $S$ and $D$
pairs \cite{FDSM2,Gi.80}.
By letting $G_0 = G_2$ and $ {\cal B}_i = b_i = 0$ for $i=1, 3$,
from Eq.\ (8, 9, 10) we have
\begin{eqnarray}
H^{even} =  E^{(e)}_0 + g_6C_{SO_6^i}  + g'_5C_{SO_5^i}
        + g'_I {\bf I}^2 ,
\label{eq27a}
\end{eqnarray}
where
\begin{eqnarray}
E^{(e)}_0 = {1\over 4} (1 - G_2) (n_1)^2 + \left[ {1\over 2}
        G_2(\Omega_1 + 6) + \varepsilon_1 \right] n_1,
\label{eq27b}
\end{eqnarray}

For odd nuclei the low-lying states are expected to belong
to the $SO_8$
irrep $[{\Omega_1\over 4} {\Omega_1\over 4} {\Omega_1\over 4}
{\Omega_1\over 4} - 1]$, corresponding to heritage $u = 1$.
The conditions for realizing the symmetry of Eq.\ (2) are
Eq.\ (\ref{eq26a}) and $g_I= 0$, which implies that
\begin{eqnarray}
B_1 - B_3 = b_1 + b_3.
\label{eq26b}
\end{eqnarray}
Under the conditions of Eqs.\ (11) and (14), and noting that for
$u = 1$ the pseudo-orbital
angular momentum is ${\bf L}^2 = 2(2+1) = 6$, we have
\begin{eqnarray}
H^{odd} =  E^{(o)}_0 + g_6C_{SO_6^i} + (g'_5 - g_5)
C_{SO_5^i} + g_5C_{SO_5^{i+k}} + g_J {\bf J}^2,
\label{eq28a}
\end{eqnarray}
\begin{eqnarray}
E^{(o)}_0 = E^{(e)}_0 + 4 g^k_5 + 6g_L + {1\over 2} G_2 (2 - \Omega_1).
\label{eq28b}
\end{eqnarray}
Using the eigenvalue formulas for the $SO_6$ and $SO_5$ Casimir operators,
the energies for even and odd systems are
\begin{eqnarray}
E^{even} = E^{(e)}_0 + g_6 \sigma (\sigma + 4) + g'_5 \tau (\tau +3)
+ g'_I J (J+1) 	,
\label{eq29a}
\end{eqnarray}
where $J(J+1)$ is used instead of $I(I+1)$ [since $L = 0$, and thus
$J = I$], and
\begin{eqnarray}
&E^{odd} =& E^{(o)}_0 + g_6 [\sigma_1 (\sigma_1 + 4) + \sigma_2
(\sigma_2 + 2) + (\sigma_3)^2] + g_J J(J+1)
\nonumber\\
&&+ (g'_5 - g_5) [\tau_1
(\tau_1 +3) + \tau_2 (\tau_2 + 1)] + g_5 [\omega_1 (\omega_1 + 3) +
\omega_2 (\omega_2 + 1)] .
\label{eq29b}
\end{eqnarray}
The reduction rules are as follows \cite{Ia.80}:
\begin{equation}
\begin{array}{rcl}
{\underline {\mbox even~ system:}} & SO_8 \supset & SO_6 \supset ~~~~
SO_5 \supset ~~~~ SO_3 \\
      & [wwww]     & [\sigma 0 0] ~~~~~~~~ [\tau 0] ~~~~~~~~~ J
\end{array}
\label{eq210}
\end{equation}
\begin{equation}
\begin{array}{rcl}
 w &=& {\Omega_1\over 4}, ~~~~ \sigma = N_1, N_1-2,..., 1 \mbox{ or } 0,\\
 \tau &=& \sigma, \sigma - 1,...,0, ~~~~ \tau = 3n_\Delta + \lambda,\\
 n_\Delta &=& 0, 1, 2,... , ~~~~ J = \lambda, \lambda +1,...,
 2\lambda -2, 2\lambda ,
\end{array}
\label{eq211}
\end{equation}
with $N_1=\frac{n_1}{2}$, where
$n_1$ is the number of the nucleons in the normal
parity levels, and
\begin{equation}
\begin{array}{rclrclr}
{\underline {\rm odd~ system:}}&( SO^i_8 \supset & SO^i_6 \supset &
SO^i_5 )
\times & SO^k_5 \supset & SO^{k+i}_5 \supset & SO^{k+i}_3 \\
  &[www,w-1] & [\sigma +{1\over 2}, {1\over 2} {1\over 2}]
  &[\tau + {1\over 2}, {1\over 2}] & [10] & [\omega_1 \omega_2] & J~~
\end{array}
\label{eq212}
\end{equation}
\begin{displaymath}
\sigma = N_1, N_1-1, ..., 1, 0; ~~~~~~ \tau = \sigma, \sigma - 1, ..., 0,
\end{displaymath}
where $N_1 = [{{n_1}\over 2}]$. The relevant Clebsch--Gordan series for $SO_5$
are \cite{Black.83}
\begin{eqnarray}
[\tau_1 {1\over 2}] \times [10] &=& [\tau_1 + 1, {1\over 2}] +
 [\tau_1 {3\over 2}] + [\tau_1 {1\over 2}] + [\tau_1 - 1, {1\over 2}],
\nonumber\\
 {[{1\over 2} {1\over 2}] } {\times} {[10]} &=& [{3\over 2} {1\over 2}]
  + [{1\over 2} {1\over 2}].
\label{eq213}
\end{eqnarray}
and the $SO_3$ content of the $SO_5$ irreps $[\omega_1 \omega_2]$ is
\cite{Ia.80}
\begin{eqnarray}
 J &=& [2(\omega_1 - \omega_2) - 6\nu_{\Delta} + {3\over 2}],
     [2(\omega_1 - \omega_2) - 6\nu_{\Delta} + {1\over 2}], ... ,
\nonumber\\
   &&  [(\omega_1 - \omega_2) - 3\nu_{\Delta} -
      {1\over 4}[1-(-)^{2\nu_{\Delta}} + {3\over 2}],
\label{eq214a}
\end{eqnarray}
\begin{eqnarray}
 \nu_{\Delta} = 0, {1\over 2}, 1, {3\over 2}, ... .
\label{eq214b}
\end{eqnarray}

The above discussions adopt as a simplification  that the numbers of
valence nucleon pairs in the normal and abnormal parity levels,
$N_1$ and $N_0$, are fixed. In reality
$N_1$ or $N_0$ have a distribution and the
semi-empirical formula of ref.\ \cite{FDSM2} has been used
to obtain $N_1$, which may generally
take non-integer values to simulate an average behavior of the
nuclear states with different values of $N_1$. For computing the
spectra and the $B(E2)$ values, we have taken the nearest integer to
the non-integer number.

The low-lying energy spectra for $^{120-132}$Xe isotopes predicted by
Eq.\ (\ref{eq29a}) are compared with data in Fig.\ 1 and
Fig.\ 2,
and the parameters used in the calculations
are given in Table I. The experimental spectra
indicate that the $SO_{3}$ parameter $g'_{I}$
is not sensitive to the neutron number in fitting the spectra of
a chain of isotopes (including both even--even and even--odd nuclei).
Therefore, in fitting the even--even nuclear spectra we fix the parameter
$g'_{I}$ to be 11.9 keV, and the adjustable parameters were taken to be
$g_{6}$ and $g'_{5}$, which will be used for the neighbouring
even-odd isotopes as well.

 From Table I, we find that $-g_{6}$ and $g'_{5}$ are nearly equal. It is
interesting to note that if the quadrupole--quadrupole interaction
is dominant over the pairing, $|B_{2}| >> |G_{0}|$  ( = $|G_{2}|$ in the
symmetry limit), and
$|B_{2}| >> |B_{3}|$ (cf.\ Eq.\ (5.3c) in ref.\ \cite{Lu.88}),
from Eq.\ (\ref{eq25b})
we obtain the
relation
\begin{eqnarray}
g'_{5} \cong -g_{6} ,
\label{eq215}
\end{eqnarray}
which is precisely the empirical relation  ${A\over 4} \cong B$ for the
parameters in the IBM $SO_6$ limit \cite{Ca.85}. The parameter $g_{6}$ can be
determined through the $0^{+}_{3} (i.e., \sigma = N_{1}-2)$ level. The good
agreement of $2^{+}_{4} (\sigma = N_{1}-2, \tau =1)$ with the experimental
results indicates that the parameters $g_{6}$ in Table I are reasonable.

Comparing Eq.\ (17) with Eqs.\ (6.2) and (6.3) of ref.\ \cite{Ba.81},
or Eqs.\ (2.9) of ref.\ \cite{Jolie.87},
we see that the spectral formulas in the $U(6/4)$ or $U(6/20)$
NSUSY and in the FDSM are essentially identical
(except for the replacement of $N$ by $N_{1}$ )
for even nuclei, while for the odd nuclei they are similar in appearance
but differ in two ways:  (1) the parameters of  the $SO(3)$ group
for the even and odd systems are different here, but the same in
$U(6/4)$ or $U(6/20)$ NSUSY. This difference comes from the
Coriolis-like  term $\bf {I\cdot L}$ in our case.  (2) There are one
$SO_{6}$ Casimir operator and two $SO_{5}$ Casimir operators for the FDSM,
in contrast to two $SO_{6}$ Casimir operators and one $SO_{5}$ Casimir
operator in the $U(6/4)$
NSUSY. These differences have a significant effect on the
spectrum. In the $U(6/4)$ NSUSY, the five lowest-energy irreps of $SO_{5}$
are $[{1\over 2} {1\over 2}], [{3\over 2} {1\over 2}],  [{5\over 2}
{1\over 2}], [{7\over 2} {1\over 2}]$ and $[{9\over 2} {1\over 2}]$; therefore
the states ${3\over 2}, {5\over 2}, {7\over 2}, {9\over 2}$ and
${11\over 2}$ belonging to the irrep $[{5\over 2} {1\over 2}]$ of $SO_{5}$
lie quite low in energy, while in the FDSM the lowest six $SO_{5}$ irreps are
$[{1\over 2} {1\over 2}]^{2}, [{3\over 2} {1\over 2}]^{2},
 [{3\over 2} {3\over 2}]$ and $[{5\over 2} {1\over 2}]$. Consequently, for the
FDSM in the low-energy region there are more low-spin states
(there are two ${1\over 2}$'s and four  ${3\over 2}$'s, while the ${7\over
2}, {9\over 2}$, and ${11\over 2}$ states are pushed up).

Using Table I and Eq.\ (\ref{eq29b}), the spectra of the
neighboring even--odd Xe
isotopes can be calculated.  For the odd-mass Xe and Ba isotopes, we
take  $g_J$ to be 35.3 keV except for $^{127}$Xe .
The difference between $g_{J}$
and $g'_{I}$ comes from the coupling of $SO^{i}_{3}$ and $SO^{k}_{3}$,
and physically is related to the Coriolis force.  We present the
calculated and experimental results for $^{127-133}$Xe in Figs.\ 3--4, and for
$^{131-135}$Ba in Fig.\ 5, with the parameters given
in Table II. Apart from the
constant term, the formula for $E^{even}$ contains three parameters, and that
for $E^{odd}$ contains two parameters beyond the three parameters
that are determined by fitting the spectra of neighboring even--even nuclei.
With two extra parameters ($g_{J}$  is kept
constant in the region discussed, except for $^{127}$Xe),
Eq.\ (\ref{eq29b}) reproduces
the spectral patterns for the nuclei $^{127-133}$Xe and $^{131-135}$Ba with
about 15 levels each.

In the $U(6/4)$ NSUSY, the ground-state
representation is $[{1\over 2} {1\over
2}]$, and thus the ground-state spin
is always ${3\over 2}$. However, experimentally
the odd nuclei in this region have both ${1\over 2}$ and ${3\over 2}$ as
the ground-state spin. The second $SO_{5}$ Casimir operator in
Eq.\ (\ref{eq29b})
provides this possibility. For alternative signs of the parameter $g_{5}$,
the ground state spin can take the values ${1\over 2}$ or ${3\over 2}$ .
What is more,
by allowing $g_{5}$ to change smoothly from positive to negative,
we can reproduce the systematic shift of the ground band of the Xe and
Ba isotopes, as shown in Fig.\ 6 and Fig.\ 7.  From Figs.\ 3--5,
we see that just as for $U(6/20)$
NSUSY \cite{Jolie.87}, for $^{129}$Xe, $^{131}$Ba  and $^{133}$Ba,
the FDSM predicts a
natural occurrence of ${1\over 2}^{+}$ as the ground state
and four low-energy ${3\over 2}^{+}$ states as the excited states.
We note that the major aim of this present FDSM description is to
give a simple and unified description of spectral
pattern of even and even--odd nuclei. For more quantitative agreement,
additional physics should be taken into account;
for instance, the mixing of particles in normal and unique parity
levels and the single-particle energy contribution.

Due to scarcity of data, we are constrained to discuss only
the
low-lying levels $(\sigma_{1}=N_{1} + {1\over 2})$.
We could in principle
compare them with levels belonging to $\sigma _{1} = N_{1} - {3\over 2}$,
If more data for higher levels were available,
This would allow us to determine the parameter
$g_{6}$ for even--odd nuclei.
Clearly, a comparison of the $g_{6}$ values obtained
from fitting even and odd nuclear spectra is a meaningful test for the
validity of the $SO_{6}$ symmetry. Here we have only
considered fitting the levels with $\tau = 1,2,3$. If the levels with $\tau >
3$
were taken into account, the $g'_{5}$ value would have to be smaller
in order to fit
the high-lying states; as a price, the unified good fit of the low-[$\tau_{1}
\tau_{2}$] states for the even--even nuclei and
even--odd nuclei would be spoiled.

Comparison of  pure $SO_{6}$ spectra with the data for
$^{120-132}$Xe (see Figs.\ 1--2) gives
reasonable agreement for $\tau \leq 3$  states for $^{120-126}$Xe.
However, the experimental energies for the high $\tau$ values in the nuclei
$^{128-132}$Xe are much lower than
predicted, with the discrepancies increasing for the higher $\tau$ values.
This
is strongly reminiscent of the stretching effect in nuclear rotational
spectra. However, a more careful comparison shows that the energy levels
within the same $SO_5$
irrep (same $\tau$) follow the $J(J+1)$ rule rather well.  Therefore the
aforementioned discrepancy cannot be due to the usual stretching
effect in which the deformation should increase with angular momentum.
In ref.\ \cite{pair}, it was pointed out that it is in fact an $SO_{5}$
$\tau$-compression effect.
The driving force for this effect is the reduction of pairing correlation with
increasing $\tau$.
Allowing for $g_{S} = G_{0} - G_{2} \not= 0$, thereby deviating from the
$SO_{6}$ limit, and treating the $g_{S}S^{\dag}\cdot S$ term as a perturbation,
leads to the following energy formula,
\begin{eqnarray}
E'^{even} \cong E^{(e)}_{0} + g_{6}\sigma (\sigma +4) +  A'\tau (\tau
+3) - B' [\tau (\tau +3)]^{2} +  g'_{I} J (J+1) .
\label{eq216}
\end{eqnarray}
Fig.\ 8 shows the spectrum for $^{126}$Xe predicted by Eqs.\ (\ref{eq29a})
and (26), respectively. Inclusion of the $SO_{5}$
stretching effect improves the agreement significantly. More examples
can be found in ref.\ \cite{pair}.

For even--odd nuclei, the levels calculated in this investigation are
limited to
those for $[\tau_{1} \tau_{1}] =[{1\over 2} {1\over 2}],[{3\over 2}
{1\over 2}]$ bands, which corresponds to the $\tau = 0,1$ states of the
even--even core. Experimental energy levels  for even--even nuclei show that
the $\tau$-compression effect is negligible for
$\tau \le 2$ states. Therefore, the
effect in the even--odd nuclei is not so conspicuous as in
the even--even nuclei.

\section {Wavefunctions}

\leftline {{\bf A. Even--even nuclei}}

According to Eq.\ (4.3a) in ref.\ \cite{Lu.88} the FDSM wavefunctions in the
$SO_{6}$ limit for $u =0$ is
\begin{eqnarray}
|N_{1}\sigma \tau n_{\Delta}IM\rangle = {\cal P}_{N_{1}\sigma \tau}
 |N_{1}\sigma \tau n_{\Delta}IM)^{IBM}_{b\rightarrow f} ,
\label{eq31a}
\end{eqnarray}
where ${\cal P}_{N_{1}\sigma \tau}$ is a Pauli
factor,
\begin{eqnarray}
{\cal P}_{N_{1}\sigma \tau} =
\left[{{(\Omega_{1} - N_{1} - \sigma)!! (\Omega_{1} - N_{1} + \sigma + 4)!!}
\over
{\Omega_{1}!! (\Omega_{1} + 4)!!}}\right]^{1\over 2} ,
\label{eq32a}
\end{eqnarray}
and $|N_{1}\sigma
\tau n_{\Delta}IM)^{IBM}_{b\rightarrow f}$ denotes a wave function
resulting from replacing the boson operators $s^{\dag}$ and $d^{\dag}_{\mu}$
by the fermion operators $S^{\dag}$ and $D^{\dag}_{\mu}$, in the
$U_{6} \supset O_{6} \supset O_{5} \supset O_{3}$ IBM wave function
$|N_{1}\sigma \tau n_{\Delta}IM)^{IBM}$,
\begin{eqnarray}
|N_{1}\sigma
\tau n_{\Delta}IM)^{IBM}_{b\rightarrow f} =
\xi_{N_{1}\sigma} (I^{\dag})^{{N_{1}-\sigma}\over 2} f_{\sigma \tau}
 (S^{\dag}, I^{\dag}) |\tau \tau \tau n_{\Delta}IM)^{IBM}_{b\rightarrow f} ,
\label{eq31b}
\end{eqnarray}
where $|\tau \tau \tau n_{\Delta}IM)^{IBM}$
denotes the IBM $U_{6} \supset U_{5} \supset SO_{5} \supset
SO_{3}$ wave function and
\begin{eqnarray}
\xi_{N_{1}\sigma} =
 \left[{{(2\sigma +4)!!}\over {(N+\sigma +4)!!(N-\sigma )!!}}
\right]^{1\over 2}~ ,
\label{eq32b}
\end{eqnarray}
\begin{eqnarray}
I^{\dag} = D^{\dag} \cdot D^{\dag} - S^{\dag} \cdot S^{\dag} ,
\label{eq33}
\end{eqnarray}
where $I^{\dag}$ is a generalized pair and $f_{\sigma \tau} (S^{\dag},
I^{\dag})$ is
a polynomial in $S^{\dag}$ and $D^{\dag}$ of order $(\sigma - \tau)$,
\begin{eqnarray}
f_{\sigma \tau} (S^{\dag},I^{\dag}) = \sum_{p=0}^{[{{\sigma - \tau}\over 2}]}
D_{p}(\sigma \tau)
(S^{\dag})^{\sigma - \tau -2p} (I^{\dag})^{p} ,
\label{eq34a}
\end{eqnarray}
\begin{eqnarray}
D_{p}(\sigma \tau) = \left[{{2^{\sigma +1}(\sigma - \tau)! (2\tau +3)!!}\over
{(\sigma +1)! (\sigma +\tau +3)!}}\right]^{1\over 2}
{{(\sigma +1 -p)!}\over {4^{p}(\sigma - \tau -2p)!p!}} .
\label{eq34b}
\end{eqnarray}
Here the notation for the wavefunctions is the same as in ref.\ \cite{Lu.88}.

\vspace{0.5cm}
\leftline {{\bf B. Even--odd nuclei}}

In this section we construct the wavefunctions for states in
odd-mass nuclei that corresponds to heritage $u = 1$. According to
the vector coherent state technique \cite{Hecht.87},
the wavefunction for the $i$-active
part can be constructed by coupling the ``collective wave function''
$|N_1 \sigma \tau n_{\Delta} I' M')^{IBM}_{b\rightarrow f}$
and the ``intrinsic state'', now the
one-fermion state $|u=1\rangle$, by means of the $SO_{6} \supset SO_{5}
\supset SO_{3}$ ~$CG$ coefficients,
\begin{eqnarray}
|2N_{1} + 1,[\sigma_{1} \sigma_{2} \sigma_{3}] [\tau_{1} \tau_{2}]
n_{\Delta}IM\rangle = K^{-1}_{l} (N_{1},[\sigma_{1} \sigma_{2} \sigma_{3}])
\left[|N_{1}[\sigma])^{IBM}_{b\rightarrow f}
|u=1\rangle\right]^{[\sigma_{1} \sigma_{2}
\sigma_{3}]}_{[\tau_{1} \tau_{2}]n_{\Delta}IM} ,
\label{eq35b}
\end{eqnarray}
where $N_{1}$ is the total number of $S, D$ pairs in the ``collective wave
function'', $l$ is the number of generalized pairs $I^{\dag},
N_{1} = \sigma + 2l$ , and $K^{-1}_{l}(N_{1},[\sigma_{1} \sigma_{2}
\sigma_{3}])$
is the diagonal matrix element of the inverse of the $K$ matrix
\cite{Hecht.87}.

Written out explicitly, Eq.\ (\ref{eq35b}) becomes
\begin{eqnarray}
&& |2N_{1}+ 1,\{\sigma_{1} + {1\over 2}, \tau_1 \} n_{\Delta}IM, [10]
k=2,m\rangle
\nonumber\\
&&~~~~ = {1\over {\sqrt {2^{N_{1}}}}} K^{-1}_{l}(N_{1},\langle \sigma +{1\over
2}\rangle)
\sum_{\tau I'n'_{\Delta}}
\xi^{\sigma \tau n'_{\Delta} I'}_{\sigma+\frac{1}{2} \tau_1 n_{\Delta} I}
\left[b^{\dag}_{2m,3/2} |N_1\sigma
\tau n'_{\Delta} I')^{IBM}_{b\rightarrow f}\right]^I_M ,
\label{eq36}
\end{eqnarray}
where the factor ${1\over {\sqrt {2^{N_{1}}}}}$ is due to the different
definitions of $A^{+}_{LM}$ in refs.\ \cite{FDSM2,Hecht.87}, the shorthand
notation stands for
\begin{eqnarray}
\left\{{\sigma +{1\over 2}},\tau_1\right\} \equiv
\left\{{\langle \sigma +{1\over 2}\rangle},
[\tau_1 {1\over 2}]\right\}, ~~~~~  {\langle \sigma +{1\over 2}\rangle} \equiv
\left[{\sigma +{1\over 2}}, {{1\over 2} {1\over 2}}\right],
\end{eqnarray}
and $\xi^{\sigma \tau n'_{\Delta} I'}_{\sigma +\frac{1}{2}
 \tau_1 n_{\Delta} I}$
is the isoscalar factor for the
group chain $SO_{6} \supset SO_{5}
\supset SO_{3}$ \cite{Ia.81}, which is a product of
the $SO_{6} \supset SO_{5}$ and $SO_{5} \supset SO_{3}$ isoscalar factors,
\begin{eqnarray}
\xi^{\sigma \tau n'_{\Delta} I'}_{\sigma+\frac{1}{2} \tau_1 n_{\Delta} I} =
\left( \begin{array}{c}
{[\sigma 0 0] \langle {1\over 2}\rangle} \\ {[\tau 0]~~~ [{1\over 2} {1\over
2}]}
\end{array}  \right| \left.
\begin{array}{c}
{\langle \sigma + {1\over 2}\rangle} \\ {[\tau_1 {1\over 2}]}
\end{array}  \right)  \left(
\begin{array}{c}
{[\tau 0] [{1\over 2} {1\over 2}]} \\ {n'_{\Delta} I'~~~ {3\over 2}}
\end{array}  \right| \left.
\begin{array}{c}
{[\tau_1 {1\over 2}]} \\ {n_{\Delta}} I
\end{array} \right) .
\label{eq37}
\end{eqnarray}
By following the steps given in ref.\ \cite{Hecht.87}, the matrix $K^{-1}_l$
is found to be
\begin{eqnarray}
K^{-1}_l(N_1,\langle \sigma + {1\over 2}\rangle) &=&\left[{{(\Omega_1 - N_1 +
\sigma + 4)!!
 (\Omega_1 - N_1 - \sigma - 2)!!} \over {2^{-N_1} (\Omega_1 - 2)!!
 (\Omega_1 + 4)!! }} \right]^{1\over 2}
\nonumber\\
& =& \left[ {{2^{-N_1} \Omega_1} \over {\Omega_1 - N_1 - \sigma}} \right]^{
 1\over 2} {\cal P}_{N_1 \sigma \tau} .
\label{eq38a}
\end{eqnarray}
Inserting (38) into (35)
\begin{eqnarray}
&& |2N_1 +1,\{\sigma + {1\over 2}, \tau_1 \} n_{\Delta} IM, [10] k=2,m \rangle
\nonumber\\
&&~~~~= \left[ {{\Omega_1} \over {\Omega_1 - N_1 + \sigma}} \right]^{
 1\over 2} \xi^{\sigma \tau n' I'}_{\sigma + {1\over 2} \tau_1 n_{\Delta} I}
 \left[ b^{\dag}_{2m {3\over 2}} | N_1 \sigma \tau n'_{\Delta} I'
\rangle \right]^I_M~~ .
\label{eq39}
\end{eqnarray}
Coupling the $i$-active and $k$-active parts gives the total wavefunction
of the $SO_6$ FDSM for even--odd nuclei
\begin{eqnarray}
&&|2N_1 +1,\{\sigma + {1\over 2},\tau_{1}\}
[\omega_1 \omega_2] n''_{\Delta} JM \rangle
\nonumber\\
&&~~~~= \sum_{I n_{\Delta}}
\left( \begin{array}{c}
{[\tau_1 {1\over 2}] [10]} \\ {n_{\Delta} I~~2}
\end{array}  \right| \left.
\begin{array}{c}
{[\omega_1 \omega_2]} \\ {n''_{\Delta} J}
\end{array}  \right)
\left[|2N_1+1,\{\sigma + {1\over 2}, \tau_1 \} n_{\Delta} I,[10]
k=2 \rangle\right]^J_M .
 \label{310}
\end{eqnarray}
Combining (34)--(40) we have
\begin{eqnarray}
&&|2N_1+1,\{\sigma + {1\over 2},\tau_{1}\}[\omega_1 \omega_2]
n''_{\Delta} JM \rangle
\nonumber\\
&&~~~~= \sum_{In_{\Delta}} \sum_{\tau I'n'_{\Delta}}
\left( \begin{array}{c}
{[\tau_1 {1\over 2}] [10]} \\ {n_{\Delta} I~~2}
\end{array}  \right| \left.
\begin{array}{c}
{[\omega_1 \omega_2]} \\ {n''_{\Delta} J}
\end{array}  \right)
\left( \begin{array}{c}
{[\sigma 0 0] \langle {1\over 2}\rangle} \\ {[\tau 0]~~ [{1\over 2} {1\over
2}]}
\end{array}  \right| \left.
\begin{array}{c}
{\langle \sigma + {1\over 2}} \\ {[\tau_1 {1\over 2}]}
\end{array} \right) \left(
\begin{array}{c}
{[\tau 0] [{1\over 2} {1\over 2}]} \\ {n'_{\Delta} I' ~~{3\over 2}}
\end{array}  \right| \left.
\begin{array}{c}
{[\tau_1 {1\over 2}]} \\ {n_{\Delta}} I
\end{array} \right)
\nonumber\\
&&~~~~~~~~\sum_j (-)^{I'+{3\over 2}+J}
\left\{ \begin{array}{ccc} {I'}&{3\over 2}&I \\ 2&J&j \end{array} \right\}
\hat I\hat J \left[ |N_1 \sigma \tau n'_{\Delta} I'\rangle a^{\dag}_j
\right]^J_M,
\label{eq311}
\end{eqnarray}
where $\hat J = \sqrt{2J+1}$.
It is interesting to note that when the fermion state for the core,
$|N_{1}\sigma \tau n'_{\Delta}I'\rangle$, is replaced by the boson state
$|N_{1}\sigma \tau n'_{\Delta}I')$
and the Pauli factor is ignored, Eq.\ (41) goes over to the
U(6/20) NSUSY wave function (2.14) of ref.\ \cite{Jolie.87}.  Thus, NSUSY can
be obtained as an approximation to the FDSM for odd-mass $SO_6$ nuclei.

\section {Electromagnetic Transitions}

\leftline{{\bf A. The $E2$ transition rate for even--even nuclei}}

In ref.\ {\cite{FDSM2}, the $E2$ transition operator in the FDSM is defined as
\begin{eqnarray}
T(E2)^{2}_{\mu} = q P^{2}_{\mu}(i)
\label{eq41}
\end{eqnarray}
while the $E2$ transition operator for the IBM $SO_{6}$ limit is {\cite{Ia.81}
\begin{eqnarray}
T(E2)^{2}_{\mu} = q B^{2}_{\mu}, ~~~~~~~
B^{2}_{\mu} = (d^{\dag} \tilde s + s^{\dag} \tilde d)^{2}_{\mu}.
\label{eq42}
\end{eqnarray}
Owing to the isomorphism between the commutators for the FDSM and IBM:
\begin{eqnarray}
 [P^{2}_{\mu}(i), S^{\dag}] \longleftrightarrow [B^{2}_{\mu}, s^{\dag}] ,
\label{eq43a}
\end{eqnarray}
\begin{eqnarray}
 [P^{2}_{\mu}(i), D^{\dag}_{\nu}] \longleftrightarrow
  [B^{2}_{\mu}, d^{\dag}_{\nu}] ,
\label{eq43b}
\end{eqnarray}
the formula for the reduced matrix elements of the $E2$ transition
operator in the FDSM is identical to that in the IBM,
\begin{eqnarray}
\langle  N\sigma \tau'n'_{\Delta}J'\parallel P^{2}(i) \parallel N\sigma \tau
n_\Delta J\rangle^{FDSM}
= (N\sigma \tau'n'_{\Delta}J'\parallel B^{2} \parallel
N\sigma \tau n_\Delta J )^{IBM}.
\label{eq44}
\end{eqnarray}
Although it is well known now, it is nevertheless a remarkable
fact that the FDSM and IBM have the same selection rules, $\Delta \sigma
= 0$ and $\Delta \tau = \pm 1$, and the same closed expression for the $E2$
transition rates \cite{IBM}. The commonly needed results are given in
Table III. It should be noted that the $O_{6}$ limit of the
IBM and the $SO_{6}$ limit of the FDSM share the same analytical form for
the spectra and the $E2$ transitions, but the accounting of the collective
pair number is different in these two model: the collective pair number is
half of the total valence nucleons ($N$) in the IBM, whereas in FDSM
it is taken as half of the total valence nucleons in the
normal parity levels ($N_{1}$).

As pointed out in ref.\ {\cite{Is.87}, when we define the $E2$ transition
operators as Eq.\ (\ref{eq41}) or Eq.\ (\ref{eq42}), the $\Delta \sigma
= 0$ and $\Delta \tau = \pm 1$ selection rules prohibit
some transitions that are observed in many nuclei. As a particular case of
these selection rules, the quadrupole moments are predicted to be
zero in the $SO_{6}$ limit, but most of the observed quadrupole moments
of the transitional nuclei differ from zero. This deviation from zero maybe
due to two causes:
one is the breaking of the $SO_{6}$ symmetry; the other
is that the $E2$ transition operator may require a more general definition.
We have chosen the later to study this problem. It is
straightforward to define a new $E2$  transition operator that relaxes
the selection rule while assuming that the wavefunctions still has
good $SO_{6} \supset SO_{5}$ symmetry. The new operator takes the form:
\begin{eqnarray}
T(E2)^{2}_{\mu} = qP^{2}_{\mu}(i) + q'(D^{\dag} \tilde D)^{2}_{\mu}.
\label{eq45}
\end{eqnarray}
The $(D^{\dag}\tilde D)^{2}_{\mu}$ term makes the following transition
possible,
\begin{eqnarray}
\Delta \sigma = \pm 2, ~~~~~~~ \Delta \tau =0, \pm 2 .
\label{eq46}
\end{eqnarray}
The reduced matrix element of $(D^{\dag} \tilde D)^{2}_{\mu}$ can be
calculated by inserting a complete set of  intermediate states (the reduced
matrix elements used here are defined according to the Rose convention)
\begin{eqnarray}
& & \langle N_{1}\sigma \tau' n'_{\Delta} L'\parallel (D^{\dag} \tilde D)^{2}
\parallel N_{1}\sigma \tau n_{\Delta} L\rangle
\nonumber\\
&&~~~~=\sqrt{5(2L+1)} \sum_{\sigma''\tau''L''} (-)^{L''+L'}
\langle N_{1}\sigma' \tau' n'_{\Delta} L'\parallel D^{\dag} \parallel
N_{1}-1 \sigma'' \tau'' n''_{\Delta} L''\rangle
\nonumber\\
&&~~~~\times \langle N_{1}\sigma \tau n_{\Delta} L \parallel D^{\dag} \parallel
N_{1}-1 \sigma'' \tau'' n''_{\Delta} L''\rangle
\left\{ \begin{array}{ccc} 2&2&2 \\ L&{L'}&{L''}
        \end{array} \right\}.
\label{eq47a}
\end{eqnarray}
The $SO_{3}$ reduced matrix element of $D^{\dag}$ in Eq.\ (49) is
\begin{eqnarray}
&& \langle N_{1} +1 \sigma' \tau' n'_{\Delta} L'\parallel D^{\dag} \parallel
N_{1}\sigma \tau n_{\Delta} L\rangle
\nonumber\\
&&~~~~= \langle N_{1}+1 \sigma' \parallel D^{\dag} \parallel N_{1}
\sigma\rangle
\left( \begin{array}{cc}
{\sigma} &  1 \\ {\tau} &  1
\end{array}  \right| \left.
\begin{array}{c}
{\sigma'} \\ {\tau'}
\end{array}  \right) \left(
\begin{array}{cc}
{\tau } &  1 \\ L &  2
\end{array}  \right| \left.
\begin{array}{c}
{\tau'} \\ {L'}
\end{array} \right),
\label{eq47b}
\end{eqnarray}
where $\langle N_{1}+1 \sigma' \parallel D^{\dag} \parallel N_{1}
\sigma\rangle$
is the $SO_{6}$ reduced matrix element and
the last two factors are the $SO_{6} \supset SO_{5}$ and $SO_{5} \supset
SO_{3}$ isoscalar factors,
respectively, which have been given in ref.\ [21] for some simple cases. The
$SO_{5}$ reduced matrix elements of $S^{\dagger}$ and $D^{\dagger}$
are given in the Appendix.

Using vector coherent state techniques {\cite{Hecht.87} for the $u=0$ case,
the $SO^{i}_{6}$
reduced matrix element of $D^{{\dag}}$ can be expressed as
\begin{eqnarray}
\langle p+1 \sigma' \parallel D^{{\dag}} \parallel p \sigma\rangle =
\sqrt {2} (\Lambda_{p+1 \sigma'} - \Lambda_{p\sigma})^{1\over 2}
\langle p+1 \sigma' \parallel z \parallel p \sigma\rangle,
\label{eq48a}
\end{eqnarray}
where $\Lambda_{p\sigma}$ is the eigenvalue of the Toronto auxiliary operator
\begin{eqnarray}
\Lambda_{p\sigma} = - {1\over 4} \left[p(p-2\Omega_{1}-6)
-\sigma(\sigma+4)\right] ,
\label{eq48c}
\end{eqnarray}
\begin{eqnarray}
\langle p+1 \sigma +1 \parallel z \parallel p\sigma\rangle =
\left[{{(\sigma+1)(l+\sigma+3)}
\over {(\sigma+3)}}\right]^{1\over 2},
\label{eq48d}
\end{eqnarray}
\begin{eqnarray}
\langle p+1 \sigma -1 \parallel z \parallel p\sigma\rangle =
\left[{{(l+1)(\sigma+3)}
\over {(\sigma+3)}}\right]^{1\over 2},
\label{eq48e}
\end{eqnarray}
with $p=N_{1}=\sigma + 2l = {1\over 2} (n_{1}-1)$. For the special cases
that will be used below, we have
\begin{eqnarray}
\langle p+1 \sigma +1 \parallel D^{\dag} \parallel p\sigma\rangle =
\left[{{(\Omega_1-2\sigma -2l)(\sigma+1)(l+\sigma+3)}
\over {(\sigma+3)}}\right]^{1\over 2} ,
\label{eq49a}
\end{eqnarray}
\begin{eqnarray}
\langle p+1 \sigma -1 \parallel D^{\dag} \parallel p\sigma\rangle =
\left[{{(\Omega_1+4-2l)(l+1)(\sigma+3)}
\over {(\sigma+1)}}\right]^{1\over 2}.
\label{eq49b}
\end{eqnarray}
In Table IV we summarize some expressions for
$B(E2)$ values and quadrupole
moments in the FDSM $SO_{6}$  limit for transitions with $\Delta \tau
=0,\pm 2$ and $\Delta \sigma = \pm 2$.
The contribution to the $B(E2)$  from the second term
$q'(D^{{\dag}} \tilde D)^{2}_{\mu}$ differs from the corresponding
term $q'(d^{{\dag}} \tilde d)^{2}_{\mu}$ in the IBM {\cite{Is.87} by the Pauli
factors.

\vspace{10pt}
\leftline{{\bf B. The transition rates for even--odd nuclei}}

For the $u=1$ case, the $E2$ transition operator can be defined as
\begin{eqnarray}
T(E2)^{2}_{\mu} = qP^{2}_{\mu}(i) + q''P^{2}_{\mu}(k).
\label{eq411}
\end{eqnarray}
The reduced matrix element is
\begin{eqnarray}
&& \langle \{N_{1}+{1\over 2},\tau'_{1}\} [\omega'_{1}\omega'_{2}]J' \parallel
qP^{2}(i) + q''P^{2}(k) \parallel
\{N_{1} + {1\over 2},\tau_{1}\} [\omega_{1}\omega_{2}]J\rangle
\nonumber\\
&&~~~~= \sum_{II'}
\left( \begin{array}{cc}
{[\tau' {1\over 2}]} &  {[10]} \\ {I'} &  2
\end{array}  \right| \left.
\begin{array}{c}
{[\omega'_{1}\omega'_{2}]} \\ {J'}
\end{array}   \right) \left(
\begin{array}{cc}
{[\tau_1 {1\over 2}]} &  {[10]} \\ I &  2
\end{array}  \right| \left.
\begin{array}{c}
{[\omega_{1}\omega_{2}]} \\ J
\end{array} \right)
M,
\label{eq412a}
\end{eqnarray}
\begin{eqnarray}
M = \langle \{N_{1} + {1\over 2},\tau'_{1}\} (I',[10]k=2)J' \parallel
qP^{2}(i) + q''P^{2}(k) \parallel
\{N_{1} + {1\over 2},\tau_{1}\} (I,[10]k=2)J\rangle .
\label{eq412b}
\end{eqnarray}
According Eq.\ (6.8) in Judd {\cite{Judd.63}}, we have
\begin{eqnarray}
[b^{{\dag}}_{ki} b_{ki}]^{K0}_{\mu 0}
 = {1\over {\sqrt{2i+1}}} \sum^n_{p=1} [b^{{\dag}}_{k}(p) b_{k}(p)]^{K}_{\mu}.
\label{eq413a}
\end{eqnarray}
Therefore, in computing the matrix elements of $P^{2}_{\mu}(k)$
the operator can be replaced by
\begin{eqnarray}
P^{2}_{\mu}(k) = \sqrt{{\Omega_1}\over 8} \sum^n_{p=1}
[b^{{\dag}}_{k}(p) b_{k}(p)]^{2}_{\mu}.
\label{eq413b}
\end{eqnarray}
Using (61) we have
\begin{eqnarray}
&M = &\hat J\hat J' (-)^{I'+J}
\left\{ \begin{array}{ccc} {J'}&J&2 \\ I&{I'}&2 \end{array} \right\}
\langle \{N_{1} + {1\over 2},\tau'_{1}\} I' \parallel qP^2(i) \parallel
\{N + {1\over 2},\tau_{1}\} I\rangle
\nonumber\\
&&+ q'' \delta_{\tau_1 \tau'_1} \delta_{II'} (-)^{I+J'} \hat J\hat J'
\sqrt{{{5\Omega_1}\over {2}}} {{\hat I}\over \hat i}
{{(\Omega_1-2N_1-1)}\over {(\Omega_1-1)}}
\left\{ \begin{array}{ccc} {J'}&J&2 \\ 2&2&I \end{array} \right\} .
\label{eq414a}
\end{eqnarray}
Now only the matrix elements of $P^{2}_{\mu}(i)$ remain to be calculated.
The generators of $Spin(6)$ for the IBFM are
\begin{eqnarray}
G^{2}_{\mu} = B^{2}_{\mu} + F^{2}_{\mu}, ~~~~~~
F^{2}_{\mu} = (a^{{\dag}}_{3\over 2} \tilde a_{3\over 2})^{2}_{\mu}
\label{eq415}
\end{eqnarray}
corresponding to the commutator for the IBFM
\begin{eqnarray}
[ F^{2}_{\mu}, a^{{\dag}}_{{3\over 2}m_i}] = (-1)^{{3\over 2}-m_i}
\langle i m_{i}+\mu,i-m_{i} | 2\mu\rangle  a^{{\dag}}_{{3\over 2}m_{i}+\mu}.
\label{eq416a}
\end{eqnarray}
There is a similar commutator in the FDSM
\begin{eqnarray}
[P^{2}_{\mu}(i), b^{{\dag}}_{22{{3\over 2}}m_{i}}] =
\sqrt{{\Omega_{1}}\over {2(2k+1)}} (-1)^{{3\over 2}-m_{i}}
\langle i m_{i}+\mu,i-m_{i} | 2\mu\rangle  b^{{\dag}}_{22{3\over 2}m_{i}+\mu},
\label{eq416b}
\end{eqnarray}
where $\sigma = \pi$  or $\nu$, and the factor $\sqrt{{\Omega_1}\over
{2(2k+1)}}$ is always equal to 1 for the 6th shell.

Because of (44, 45) and (64, 65), we have the following isomorphism between
the commutators in the FDSM and IBFM,
\begin{eqnarray}
[P^{2}_{\mu}, S^{{\dag}}] \longleftrightarrow [B^{2}_{\mu}, s^{{\dag}}]
= [G^{2}_{\mu}, s^{{\dag}}],
\label{eq417a}
\end{eqnarray}
\begin{eqnarray}
[P^{2}_{\mu}(i), D^{{\dag}}_{\nu}] \longleftrightarrow [B^{2}_{\mu},
d^{{\dag}}_{\nu}]
= [G^{2}_{\mu}, d^{{\dag}}_{\nu}],
\label{eq417b}
\end{eqnarray}
\begin{eqnarray}
[P^{2}_{\mu}(i), b^{{\dag}}_{22{3\over 2}m_{i}}] \longleftrightarrow
[F^{2}_{\mu}, a^{{\dag}}_{{3\over 2}m_{i}}]
 = [G^{2}_{\mu}, a^{{\dag}}_{{3\over 2}m_{i}}] .
\label{eq417c}
\end{eqnarray}
Therefore we establish the following identity:
\begin{eqnarray}
&&\langle \{N+{1\over 2},\tau'_{1}\}I' \parallel P^{2}(i) \parallel
\{N+{1\over 2},\tau_{1}\}I \rangle ^{FDSM}
\nonumber\\
&&~~~~~~ = \left(\{N+{1\over 2},\tau'_{1}\}I'
\parallel G^{2} \parallel \{N+{1\over 2}, \tau_{1}\}I\right)^{IBFM} .
\label{eq418}
\end{eqnarray}
The reduced matrix element of $G^{2}_{\mu}$ is derived in ref.\ {\cite{Ia.81}.
With these results we can calculate the $B(E2)$ values and the quadrupole
moments for odd-mass nuclei.
The selection rules for U(6/4) are {\cite{Ia.81}
\begin{eqnarray}
\Delta \tau_{1} = 0,\pm 1, ~~~~~~ \Delta \tau_{2} = 0.
\label{eq419}
\end{eqnarray}
For the $u=1$ case in the FDSM, owing to the Kronecker product (22)
the corresponding selection rules are
\begin{eqnarray}
\Delta \omega_{1} = 0,\pm 1, \pm 2, \pm 3, ~~~~~~
\Delta \omega_2 = 0, \pm 1,
\label{eq420}
\end{eqnarray}
With these rules, the restrictions for the $B(E2)$ values will be less
severe that of the IBFM {\cite{Ia.81}. This enables us to explain some
data that cannot be explained by the U(6/4) IBFM.

 From Table III and Table IV, we can calculate the $B(E2)$ values
for the transitions $\Delta \omega_{1} = 0,\pm 1, \pm 2$.
In order to make a direct
comparison between the calculated and experimental results without
a knowledge of $q$ and $q'$, we compute the relative $B(E2)$ value
rather than the absolute values. The FDSM prediction and the
experimental results {\cite{Ca.85,Re.88} for Xe and Ba isotopes are listed
in Tables V and VI. For the transitions with $\Delta \sigma = 0$ and
$\Delta \tau = \pm 1$, the formulas for the $B(E2)$ in the IBM and FDSM
are of the
same form, but the numerical values differ for a given nucleus because
in the IBM the $B(E2)$ is a function of $N$, while  in the
FDSM it  is a function of $N_{1}$. For the transitions with
$\Delta \sigma = \pm 2$ and $\Delta \tau = 0, \pm 2$,
they also
differ by the Pauli factors $(\Omega_{1}-2N_{1}+2)^{2}$ or $(\Omega_{1}+4)
(\Omega_{1}-2N_{1}+2)$, as shown in Table IV.

 From Tables V and VI, we can see that the $B(E2)$ transitions for
Xe and Ba isotopes exhibit an $SO_{6}$ symmetry, especially
for the $\Delta \tau = \pm 1$ transitions. There are two reasons to expect
less accuracy for the $\Delta \tau = \pm 2$ transitions: one is the
definition of the new $T(E2)$ operator and the other is
the fitting of the parameter $[{q'\over q}]^{2}$. In fact, the
determination of $[{q'\over q}]^{2}$ from the rate $B(E2, 2^{+}_{2}
\rightarrow 0^{+}_{1})/B(E2,2^{+}_{2} \rightarrow 2^{+}_{1})$
is very inaccurate. A possible way to
obtain $q$ and $q'$ is, as in ref.\ [17], through fitting the
$B(E2,2^+_{1} \rightarrow
0^{+}_{1}$) and the quadrupole moment $Q(2^{+}_{1}$), respectively.
For example, from $Q(2^{+}_{1})=-0.16$ eb and $B(E2,2^{+}_{1} \rightarrow
0^{+}_{1}$) = 0.146 (eb)$^{2}$ for $^{134}$Ba, we can determine
$({q'\over q})^{2}$  to be equal to 0.34, which in turn gives the $B(E2)$
value listed in the last column (theo.b) of Table VI.
By comparing the last
two columns in Table VI, we see that the last column gives a better
fit. If more $Q(2^{+}_{1}$) values were available, it
would be  possible to
obtain a better description of the relative $B(E2)$ for the $\Delta \tau = \pm
2$ transitions.

In ref.\ \cite{Br.88}
the ratio $R_{4}$ between two $B(E2)$ values is introduced to
distinguish the $SO_{6}$ limit from the $U_{5}$  limit of the IBM,
\begin{eqnarray}
R_{4} = {{B(E2, 4^{+}_{1} \rightarrow 2^{+}_{1})}\over
{B(E2, 4^{+}_{1} \rightarrow 0^{+}_{1})}} .
\label{eq421}
\end{eqnarray}
The explicit expression for R$_{4}$  predicted by the IBM is
\begin{eqnarray}
R_4 = \left\{ \begin{array}{ll}
\displaystyle{{2(N-1)}\over N} & \mbox{for the $U_5$ limit}, \\[5pt]
\displaystyle{{10(N-1)(N+5)}\over {7N(N+4)}} & \mbox{for the $SO_6$ limit}
\end{array} \right. ,
\label{eq422}
\end{eqnarray}
where $N$  is the boson number. The R$_{4}$  value derived from
the FDSM has
the same form as above, but with $N$ replaced by $N_{1}$,
\begin{eqnarray}
R_{4} = {{10(N_{1}-1)(N_{1}+5)}\over {7N_{1}(N_{1}+4)}}
\label{eq423}
\end{eqnarray}
The $N_{1}$  values can be estimated from shell model
configurations of protons and neutrons in the odd-A nuclei, and are
shown in the Table 7.1 of ref.\ \cite{Pr.75}.
In Table VII
we list the FDSM prediction for R$_{4}$  along with the experimental
results of ref.\ \cite{Br.88}.
It can be seen that the SO$_{6}$  limit of the FDSM seems
to explain the experimental data better than the IBM.
Alternatively, we note that if accurate R$_{4}$ values are
available, we may be used  to
obtain the  empirical $N_{1}$ value from Eq.\ (74).

It should be mentioned again that apart from the Pauli effect,
the FDSM differs from the IBM \cite{Re.88}
in the value of the number of the
collective pairs ($N_1$ vs.\ $N$).
The spectrum of the
$\sigma = N_{1}$ band is not sensitive to the value of $N_{1}$,
but the observation
that the parameters $g_{6}$  and $g'_{5}$ in Table I change smoothly
between nuclei, and that the experimental spectra for the even and odd
nuclei can be fitted with the same $g_{6}$  and $g'_{5}$
values, suggest that
the choice of $N_{1}$  taken in this paper is reasonable.

The difference between $N$ and $N_1$ does affect the energies
for the bands with $\sigma = N_1-2, N_1-4 \ldots$
In ref.\ \cite{Ca.85}, it is pointed out that
the parameters for the $SO_6$ nuclei in both the A=130 and Pt regions
have a common characteristic that $g'_5 \cong -g_6$ (see
Eq.\ (28)). With such an empirical relation, the following energy
ratio has a simple form in both the IBM and the FDSM
\begin{eqnarray}
{{E^{+}_{O_{3}} (\sigma =N-2)}\over {E^{+}_{2_{2}} - E^{+}_{2_{1}}}}
 = {2(N+1)\over 3},
\label{eq424a}
\end{eqnarray}
\begin{eqnarray}
{{E^{+}_{O_{3}}(\sigma =N_{1}-2)}\over {E^{+}_{2_{2}}-E^{+}_{2_{1}}}}
 = {2(N_{1}+1)\over 3},
\label{eq424b}
\end{eqnarray}

A comparison of the ratios calculated using Eqs.\ (75, 76) and the
experimental data is shown in Table VIII.
This example indicates also that the FDSM $SO_{6}$ model reproduces this ratio
better than the IBM $SO_{6}$ for nuclei in the Xe--Ba region.
This suggests that in this region  an empirical effective boson number maybe
needed to give a better agreement with data in IBM calculations.

The $E2$  transition rate is  generally more sensitive
to the parameter $N_{1}$ than the spectra. The reasonableness of the
chosen $N_{1}$ value can also be seen from the good agreement
between the calculated and experimental values of the $B(E2)$ values for the
isotopes of Ba  and Xe,
as shown in Table V and  VI.  Finally, we reiterate that
as $N_{1}$ increases in the shell, the spins for the ground states
of odd nuclei swap naturally between ${1\over 2}$ and ${3\over 2}$ (this
transition occurs
at $^{131}$Xe and $^{135}$Ba for the isotopes of Xe and Ba, respectively).

In Table IX, both experimental and theoretical
 $B(E2)$ values for the even--odd nuclei of $^{129}$Xe and
$^{131}$Xe are given, and compared with the calculated results of
the NSUSY case.
Here the effective charges (i.e., $q$) are the same as the neighbouring
even-even nuclei,
and determined by the experimental $B(E2,2^{+}_{1} \rightarrow
0^{+}_{1}$) values. While effective charges for k-active part (i.e., $q''$)
are fitted by E2 transitions of even--odd nuclei.
In this work, ($q$, $q''$)=(0.129,0.075) $eb$, (0.143,0.093) $eb$ for
$^{129}$Xe and $^{131}$Xe respectively.
The agreement with data is comparable in the two cases, although the NSUSY
calculations give a somewhat better agreement of the weaker transitions.
Finally, we reiterate that
the group chain (2) is very similar to the
NSUSY group chain (\ref{eq11}).  However, the pseudo orbital angular
momentum 2 is introduced {\it ad hoc} in the NSUSY, while in the FDSM
it is a natural result of the reclassification (in terms of the
$k$-$i$ basis) of the shell model single-particle states for the sixth shell.

\section{Conclusions}

In this work, we provide simple but unified analytic solutions
of even and even--odd nuclei within the framework of the fermion dynamical
symmetry model.
The good agreement of both level pattern and E2 transitions with our simplified
solutions indicates
a good SO(6) symmetry for both even and even-odd nuclei in A=130 region.
We find that generally the FDSM results provide a unified description of the
even and odd nuclei in this region that is comparable to or even somewhat
better than IBM and NSUSY approaches, but to a lesser degree in
phenomenology.

{\bf Acknowledgement}

This work was supported by the NSF (Drexel). The research in CYCU
was supported by the National Science Council of ROC. Theoretical
nuclear physics research at the University of Tennessee is supported
by the U.~S. Department of Energy through Contracts No.\
DE--FG05--93ER40770 and DE-FG05-87ER40461.  Oak Ridge National
Laboratory is managed by Lockheed Martin Energy Systems, Inc.\ for
the U.~S. Department of Energy under Contract No.\ DE--AC05--84OR21400.
We thank Prof.\ J.\ Q.\ Chen for stimulating discussions and co-operation.
X.\ W.\ P.\ and D.\ H.\ F.\ are grateful to Dr.\ Sun Yang for typesetting
part of the manuscript.

\baselineskip = 14pt
\bibliographystyle{unsrt}

\newpage

\begin{center}

Table I.\ Parameters for the even Xe isotopes.

\begin{tabular}{cccc} \hline
  Nuclei  &  $ g_6$(keV) &
 $ g'_5$(keV) & $ g'_I$(keV) \\ \hline
 $ ^{120}$Xe &  -60   &  53  &  11.9 \\
 $ ^{122}$Xe &  -64   &  59  &  11.9 \\
 $ ^{124}$Xe &  -68.8 &  64  &  11.9 \\
 $ ^{126}$Xe &  -73.3 &  71  &  11.9 \\
 $ ^{128}$Xe &  -78.2 &  79  &  11.9 \\
 $ ^{130}$Xe &  -100.9 & 100  &  11.9 \\
 $ ^{132}$Xe &  -106.5 & 122  &  11.9 \\ \hline
\end{tabular}

\vspace{0.2in}

Table II-a.\ Parameters for the odd Xe isotopes.

\begin{tabular}{ccccc} \hline
  Nuclei  & $ g_6$(keV) &
 $ g'_5$(keV) & $ g_5$(keV) & $g_J$(keV) \\ \hline
 $ ^{127}$Xe & -73.3  &  71.0  & -38.0 & 25.0 \\
 $ ^{129}$Xe & -78.2  &  79.0  & -18.0 & 35.3 \\
 $ ^{131}$Xe & -100.9 & 100.0  &  30.0 & 35.3 \\
 $ ^{133}$Xe & -106.5 & 122.0  &  50.5 & 35.3 \\
 $ ^{135}$Xe &        & 142.1  &  70.6 & 35.3 \\ \hline
\end{tabular}

\vspace{0.2in}

Table II-b.\ Parameters for the odd Ba isotopes.

\begin{tabular}{cccc} \hline
  Nuclei  & $ g^i_5$(keV) & $ g_5$(keV) & $g_J$(keV) \\ \hline
 $ ^{131}$Ba & 72.5  &  -20  &  35.3 \\
 $ ^{133}$Ba & 105 &  -15  &  35.3 \\
 $ ^{135}$Ba & 90  &   50  &  35.3 \\
 $ ^{137}$Ba & 72  &   70  &  35.3 \\ \hline
\end{tabular}

\newpage

Table III.\ $B(E2)$ formulas for even--even nuclei in the $SO_6$ limit.

\begin{tabular}{lcll} \hline
$|N_1\sigma \tau J_i \rangle $ & $ \rightarrow $ &
 $|N_1\sigma' \tau' J_f \rangle $ & $ B(E2;J_i \rightarrow J_f)$ \\ \hline
\vspace*{0.1in}
$|N_1 N_1~1+L/2~L+2 \rangle $ & $ \rightarrow $ &
 $|N_1 N_1~L/2~L \rangle $ &
 $ q^2 \displaystyle{\frac{(L+2)(2N_1+L+8)}{8(L+5)}}(2N_1-L)$ \\
\vspace*{0.1in}
$|N_1 N_1~1~2 \rangle $ & $ \rightarrow $ &
 $|N_1 N_1~0~0 \rangle $ & $ \displaystyle{\frac{1}{5}}q^2 N_1(N_1+4) $ \\
\vspace*{0.1in}
$|N_1 N_1~2~2 \rangle $ & $ \rightarrow $ &
 $|N_1 N_1~1~2 \rangle $ & $ \displaystyle{\frac{2}{7}}q^2 (N_1-1)(N_1+5) $ \\
\vspace*{0.1in}
$|N_1 N_1~3~4 \rangle $ & $ \rightarrow $ &
 $|N_1 N_1~2~4 \rangle $ & $ \displaystyle{\frac{10}{63}}q^2 (N_1-2)(N_1+6) $
\\
\vspace*{0.1in}
$|N_1 N_1~3~3 \rangle $ & $ \rightarrow $ &
 $|N_1 N_1~2~4 \rangle $ & $ \displaystyle{\frac{2}{21}}q^2 (N_1-2)(N_1+6) $ \\
\vspace*{0.1in}
$|N_1 N_1~3~3 \rangle $ & $ \rightarrow $ &
 $|N_1 N_1~2~2 \rangle $ & $ \displaystyle{\frac{5}{21}}q^2 (N_1-2)(N_1+6) $ \\
\vspace*{0.1in}
$|N_1 N_1~3~4 \rangle $ & $ \rightarrow $ &
 $|N_1 N_1~2~2 \rangle $ & $ \displaystyle{\frac{11}{63}}q^2 (N_1-2)(N_1+6) $
\\
\vspace*{0.1in}
$|N_1 N_1~3~0 \rangle $ & $ \rightarrow $ &
 $|N_1 N_1~2~2 \rangle $ & $ \displaystyle{\frac{7}{21}}q^2 (N_1-2)(N_1+6) $ \\
\hline
\end{tabular}

\newpage

Table IV.\ Some quadrupole moments and $B(E2)$ values
for the transitions $\Delta \sigma=0,\pm 2, \Delta \tau=0, \pm 2$.

\begin{tabular}{ll} \hline
\vspace*{0.1in}
$B(E2;N_1 N_1~2~2 \rightarrow  N_1 N_1~0~0 )$  &
 $= q^{\prime 2} \displaystyle{\frac{N_1(N_1-1)(N_1+4)(N_1+5)}
 {70(N_1+1)^2}(\Omega_1-2N_1+2)^2} $ \\
\vspace*{0.1in}
$B(E2;N_1 N_1~3~3 \rightarrow  N_1 N_1~1~2 )$  &
 $= q^{\prime 2} \displaystyle{\frac{5(N_1-1)(N_1-2)(N_1+5)(N_1+6)}
 {294(N_1+1)^2}(\Omega_1-2N_1+2)^2} $ \\
\vspace*{0.1in}
$B(E2;N_1 N_1~3~4 \rightarrow  N_1 N_1~1~2 )$  &
 $= q^{\prime 2} \displaystyle{\frac{11(N_1-1)(N_1-2)(N_1+5)(N_1+6)}
 {882(N_1+1)^2}(\Omega_1-2N_1+2)^2} $ \\
\vspace*{0.1in}
$B(E2;N_1 N_1~3~4 \rightarrow  N_1 N_1~3~3 )$  &
 $= q^{\prime 2} \displaystyle{\frac{2(N_1^2+4N_1+23)^2}
 {231(N_1+1)^2}(\Omega_1-2N_1+2)^2} $ \\
\vspace*{0.1in}
$B(E2;N_1 N_1~3~0 \rightarrow  N_1 N_1~1~2 )$  &
 $= q^{\prime 2} \displaystyle{\frac{(N_1-1)(N_1-2)(N_1+5)(N_1+6)}
 {42(N_1+1)^2}(\Omega_1-2N_1+2)^2} $ \\
\vspace*{0.1in}
$B(E2;N_1 N_1~2~2 \rightarrow  N_1 N_1~2~4 )$  &
 $= q^{\prime 2} \displaystyle{\frac{16(N_1^2+4N_1+15)^2}
 {2205(N_1+1)^2}(\Omega_1-2N_1+2)^2} $ \\
\vspace*{0.1in}
$B(E2;N_1 N_1-2~0~0 \rightarrow  N_1 N_1~1~2 )$  &
 $= q^{\prime 2} \displaystyle{\frac{(N_1+2)(N_1+3)(N_1+4)(N_1+5)}
 {14N_1(N_1+1)^2}(\Omega_1-2N_1+2)(\Omega_1+4)} $ \\
\vspace*{0.1in}
$B(E2;N_1 N_1-2~1~2 \rightarrow  N_1 N_1~1~2 )$  &
 $= q^{\prime 2} \displaystyle{\frac{(N_1-1)(N_1-2)(N_1+3)(N_1+4)}
 {49N_1(N_1+1)^2}(\Omega_1-2N_1+2)(\Omega_1+4)} $ \\
\vspace*{0.1in}
$Q(N_1 N_1~1~2) $ &
 $= q^{\prime} \displaystyle{\sqrt{\frac{2\pi}{35}}\frac{4(N_1^2+4N_1+9)}
 {7(N_1+1)}(\Omega_1-2N_1+2)} $ \\ \hline
\end{tabular}

\newpage

Table V.\ Relative $B(E2)$ values for the even Xe isotopes.

\begin{tabular}{rllllllllll} \hline
    & \multicolumn{2}{c}{$^{120}$Xe}
    & \multicolumn{2}{c}{$^{124}$Xe}
    & \multicolumn{2}{c}{$^{126}$Xe}
    & \multicolumn{2}{c}{$^{128}$Xe}
    & \multicolumn{2}{c}{$^{130}$Xe} \\
 $ J_i \rightarrow J_f $ & Exp. & Theo. &  Exp. & Theo.
 &  Exp. & Theo. &  Exp. & Theo. &  Exp. & Theo. \\ \hline
 $ 2^+_2 \rightarrow 2^+_1 $ & 100  & 100 & 100 & 100
 & 100 & 100 & 100 & 100 & 100 & 100  \\
 $       \rightarrow 0^+_1 $ & 5.6  & 5.6 & 3.9 & 3.9
 & 1.5 & 1.4 & 1.2 & 1.2 & 0.6 & 0.6  \\
 $ 3^+_1 \rightarrow 2^+_2 $ & 100  & 100 & 100 & 100
 & 100 & 100 & 100 & 100 & 100 & 100  \\
 $       \rightarrow 4^+_1 $ & 50   & 40  & 46  & 40
 & 34  & 40  & 37  & 40  & 25  & 40   \\
 $       \rightarrow 2^+_1 $ & 2.7  & 7.1 & 1.6 & 4.9
 & 2.0 & 1.85 & 1.0 & 1.5 & 1.4 & 0.72  \\
 $ 4^+_2 \rightarrow 2^+_2 $ & 100  & 100 & 100 & 100
 & 100 & 100 & 100 & 100 & 100 & 100  \\
 $       \rightarrow 4^+_1 $ & 62   & 91  & 91  & 91
 & 76  & 91  & 133 & 91  & 107 & 91   \\
 $       \rightarrow 2^+_1 $ & --- & 7.11 & 0.4 & 4.91
 & 1.0 & 1.83 & 1.7 & 1.49 & 3.2 & 0.97  \\
 $ 0^+_2 \rightarrow 2^+_2 $ & 100  & 100 & 100 & 100
 & 100 & 100 & 100 & 100 & 100 & 100  \\
 $       \rightarrow 2^+_1 $ & --- & 7.11 & 1   & 4.91
 & 7.7 & 1.83 & 14  & 1.49 & 26  & 0.97  \\ \hline
\end{tabular}
\end{center}
\vspace{-6mm}

\newpage

\begin{center}

Table VI.\ Relative $B(E2)$ values for the even Ba isotopes.
\begin{tabular}{rlllllllllll} \hline
    & \multicolumn{2}{c}{$^{126}$Ba}
    & \multicolumn{2}{c}{$^{128}$Ba}
    & \multicolumn{2}{c}{$^{130}$Ba}
    & \multicolumn{2}{c}{$^{132}$Ba}
    & \multicolumn{2}{c}{$^{134}$Ba} \\
 $ J_i \rightarrow J_f $ & Exp. & Theo. &  Exp. & Theo.
 &  Exp. & Theo. &  Exp. & Theo. &  Exp. & Theo.a & Theo.b \\ \hline
 $ 2^+_2 \rightarrow 2^+_1 $ & 100  & 100 & 100 & 100
 & 100 & 100 & 100 & 100 & 100 & 100 & 100 \\
 $       \rightarrow 0^+_1 $ & 11  & 11 & 9.2 & 9.2
 & 5.7 & 5.7 & 0.2 & 0.2 & 1.0 & 1.1 & 3.06  \\
 $ 3^+_1 \rightarrow 2^+_2 $ & 100  & 100 & 100 & 100
 & 100 & 100 & 100 & 100 & 100 & 100 & 100 \\
 $       \rightarrow 4^+_1 $ & 13   & 40  & --- & 40
 & 30  & 40  & 73  & 40  & 40  & 40 & 40  \\
 $       \rightarrow 2^+_1 $ & 5.8  & 14.4 & --- & 12
 & 1.5 & 7.46 & 0.2 & 0.24 & 1.0 & 1.26 & 3.89  \\
 $ 4^+_2 \rightarrow 2^+_2 $ & 100  & 100 & 100 & 100
 & 100 & 100 & 100 & 100 & 100 & 100 & 100 \\
 $       \rightarrow 3^+_1 $ & ---  & 26.1 & --- & 21.8
 & --- & 13.5 & --- & 0.44 & 14.5 & 3.06 & 9.45  \\
 $       \rightarrow 4^+_1 $ & 28   & 91  & 42  & 91
 & 89  & 91  & 75 & 91  & 77 & 91 & 91  \\
 $       \rightarrow 2^+_1 $ & 1.1  & 14.4 & 1.7 & 12
 & 3.9 & 7.46 & 2.2 & 0.24 & 2.5 & 1.26 & 3.89  \\
 $ 0^+_2 \rightarrow 2^+_2 $ & 100  & 100 & 100 & 100
 & 100 & 100 & 100 & 100 & 100 & 100 & 100 \\
 $       \rightarrow 2^+_1 $ & --- & 14.4 & --- & 12
 & --- & 7.46 & 0    & 0.24 & 4 & 1.26  & 3.89  \\ \hline
\end{tabular}
\end{center}
\vspace{-6mm}

\vspace{0.2in}

\begin{center}

Table VII.\ The value of $R_4$.

\begin{tabular}{ccccccc} \hline
 $ Nuclei $ & $N$ & $N_1$ & $R^{exp}_4$ &
 $R^{FDSM}_4(SO_6)$ & $R^{IBM}_4(SO_6)$ & $R^{IBM}_4(U_5)$ \\ \hline
 $ ^{120}$Xe & 10 & 7 & 1.46(20) & 1.34 &  1.38 & 1.80 \\
 $ ^{124}$Xe &  8 & 6 & 1.29(15) & 1.31 &  1.35 & 1.75 \\
 $ ^{126}$Ba &  9 & 6 & 1.12(20) & 1.34 &  1.37 & 1.75 \\
 $ ^{128}$Ba &  8 & 6 & 1.03(14) & 1.31 &  1.35 & 1.75 \\
 $ ^{130}$Ba &  7 & 5 & 0.90(13) & 1.27 &  1.34 & 1.71 \\
 $ ^{130}$Xe &  5 & 4 & 1.35(18) & 1.21 &  1.27 & 1.60 \\ \hline
\end{tabular}

\newpage

Table VIII.\ The $E(0^+_3)/[E(2^+_2)-E(2^+_1)]$ ratio.

\begin{tabular}{cccccc} \hline
 $ Nuclei $ & $N$ & $N_1$ & Exp. & IBM & FDSM  \\ \hline
 $ ^{118}$Xe & 9 & 6 & 2.912 & 6.67 & 4.67  \\
 $ ^{122}$Xe & 9 & 6 & 3.68  & 6.67 & 4.67  \\
 $ ^{124}$Xe & 8 & 6 & 3.44  & 6.00 & 4.67  \\
 $ ^{126}$Xe & 7 & 5 & 3.58  & 5.33 & 4.00  \\
 $ ^{128}$Xe & 6 & 5 & 3.56  & 4.67 & 4.00  \\
 $ ^{130}$Xe & 5 & 4 & 3.44  & 4.00 & 3.33  \\
 $ ^{134}$Ba & 5 & 5 & 3.84  & 4.00 & 4.00  \\
 $ ^{136}$Ba & 4 & 4 & 2.918 & 3.33 & 3.33  \\
 $ ^{138}$Ce & 5 & 5 & 3.25  & 4.00 & 4.00  \\ \hline
\end{tabular}

\newpage

Table IX.\ Transition probabilities in $^{129}$Xe ($N_1=5$) and
 $^{131}$Xe ($N_1$=4), $\Omega_1=20$.

\begin{tabular}{clll|clll} \hline
 \multicolumn{4}{c|}{$^{129}$Xe} &
 \multicolumn{4}{c}{$^{131}$Xe} \\ \hline
   & \multicolumn{3}{c|}{$B(E2) (e^2b^2)$} &
   & \multicolumn{3}{c}{$B(E2) (e^2b^2)$} \\ \cline{2-4} \cline{6-8}
 $ J_i \rightarrow J_f $ & FDSM  & ~~Exp. & Ref.[5] &
 $ J_i \rightarrow J_f $ & FDSM  & ~~Exp. & Ref.[5] \\ \hline
 $ \frac{3}{2}^+_1 \rightarrow \frac{1}{2}^+_1 $
 & 0.036 &      & 0.007   &
 $ \frac{1}{2}^+_1 \rightarrow \frac{3}{2}^+_1 $
 & 0.0953  & ~~0.0039 & 0.0012 \\

 $ \frac{3}{2}^+_2 \rightarrow \frac{3}{2}^+_1 $
 & 0.0186  & $<$0.0005 & 0.013 &
 $ \frac{5}{2}^+_1 \rightarrow \frac{1}{2}^+_1 $
 & 0.075  & ~~0.030 & 0.016 \\

 $ \frac{3}{2}^+_2 \rightarrow \frac{1}{2}^+_1 $
 & 0.084  & ~~0.12 & 0.12 &
 $ \frac{5}{2}^+_1 \rightarrow \frac{3}{2}^+_1 $
 & 0.004  & ~~0.10 & 0.10  \\

 $ \frac{5}{2}^+_1 \rightarrow \frac{3}{2}^+_1 $
 & 0.011  & ~~0.22 & 0.10 &
 $ \frac{3}{2}^+_2 \rightarrow \frac{3}{2}^+_1 $
 & 0.053  & ~~0.057 & 0.058 \\

 $ \frac{5}{2}^+_1 \rightarrow \frac{1}{2}^+_1 $
 & 0.070  & ~~0.077 & 0.039 &
 $ \frac{1}{2}^+_2 \rightarrow \frac{1}{2}^+_1 $
 & 0.0000  &          \\

 $ \frac{1}{2}^+_2 \rightarrow \frac{3}{2}^+_2 $
 & 0.028  &     &    &
 $ \frac{1}{2}^+_2 \rightarrow \frac{3}{2}^+_1 $
 & 0.124  & ~~0.048 & 0.115 \\

 $ \frac{1}{2}^+_2 \rightarrow \frac{3}{2}^+_1 $
 & 0.14  & ~~0.044 & 0.12 &
 $ \frac{7}{2}^+_1 \rightarrow \frac{5}{2}^+_1 $
 & 0.028  & ~~0.005 & 0.0013 \\

 $ \frac{1}{2}^+_2 \rightarrow \frac{1}{2}^+_1 $
 & 0.0000  &   &      &
 $ \frac{7}{2}^+_1 \rightarrow \frac{3}{2}^+_1 $
 & 0.043  & ~~0.081 & 0.082  \\

 $ \frac{5}{2}^+_2 \rightarrow \frac{1}{2}^+_1 $
 & 0.004  & ~~0.057 & 0.071 &
 $ \frac{3}{2}^+_3 \rightarrow \frac{3}{2}^+_1 $
 & 0.025  & ~~0.027 & 0.017 \\

 $ \frac{3}{2}^+_3 \rightarrow \frac{1}{2}^+_1 $
 & 0.056  & ~~0.0032 & 0.0056 &
 $ \frac{5}{2}^+_2 \rightarrow \frac{3}{2}^+_2 $
 & 0.004  & $<$0.031 & 0.0011 \\

 $ \frac{3}{2}^+_4 \rightarrow \frac{1}{2}^+_1 $
 & 0.0133  & ~~0.0030 & 0.0004 &
 $ \frac{5}{2}^+_2 \rightarrow \frac{1}{2}^+_1 $
 & 0.071  & ~~0.068 & 0.056  \\

   &   &   &   &
 $ \frac{5}{2}^+_2 \rightarrow \frac{3}{2}^+_1 $
 & 0.014  & ~~0.013 & 0.043 \\
   &   &   &   &
 $ \frac{7}{2}^+_2 \rightarrow \frac{3}{2}^+_1 $
 & 0.124  & ~~0.005 & 0.026 \\ \hline
\end{tabular}

\end{center}

\newpage

\begin{center}
{\bf Appendices }

\vspace{0.2in}

 The $SO_5$ reduced matrix elements
\end{center}

\renewcommand{\theequation}{A-\arabic{equation}}
\setcounter{equation}{0}

\begin{equation}
\langle N+1,\sigma +1,\tau \parallel S^{\dag} \parallel N \sigma \tau\rangle
=
\left[{{(\Omega_{1}-\sigma-N)(\sigma-\tau+1)(\sigma+\tau+4)(N+\sigma+6)}\over
{4(\sigma+2)(\sigma+3)}}\right]^{1\over 2}
\end{equation}

\begin{equation}
\langle N+1,\sigma -1,\tau \parallel S^{\dag} \parallel N \sigma \tau\rangle
=
-\left[{{(\Omega_{1}+\sigma-N+4)(\sigma-\tau)(\sigma+\tau+3)(N-\sigma+2)}\over
{4(\sigma+1)(\sigma+2)}}\right]^{1\over 2}
\end{equation}

\begin{eqnarray}
&& \langle N+1,\sigma +1,\tau' \parallel D^{\dag} \parallel N \sigma
\tau\rangle  =
\nonumber \\
&& ~~~\left[ \frac{(\Omega_1-\sigma-N)(N+\sigma+6)}{4(\sigma+2)(\sigma+3)}
 \right]^{\frac{1}{2}} \times
 \left\{
 \begin{array}{cc}
 \left[ \displaystyle{\frac{(\sigma+\tau+4)(\sigma+\tau+5)(\tau+1)}
 {(2\tau+5)}} \right]^{1\over 2}, & \tau'=\tau+1 \\
 -\left[ \displaystyle{\frac{(\sigma-\tau+1)(\sigma-\tau+2)(\tau+2)}
 {(2\tau+1)}} \right]^{{1\over 2}}, & \tau'=\tau-1 \end{array} \right.
\end{eqnarray}

\begin{eqnarray}
&& \langle N+1,\sigma -1,\tau' \parallel D^{\dag} \parallel N \sigma
\tau\rangle  =
\nonumber \\
&& ~~~\left[ \frac{(\Omega_{1}+\sigma-N+4)(N-\sigma+2)}{4(\sigma+1)(\sigma+2)}
 \right]^{\frac{1}{2}} \times
 \left\{ \begin{array}{cc} -\left[
 \displaystyle{\frac{(\sigma+\tau+2)(\sigma+\tau+3)
 (\tau+2)}{(2\tau+1)}} \right]^{{1\over 2}}, & \tau'=\tau+1 \\
 \left[ \displaystyle{\frac{(\sigma-\tau-1)(\sigma-\tau)(\tau+1)}{(2\tau+5)} }
 \right]^{{1\over 2}},
 & \tau'=\tau-1 \end{array} \right.
\end{eqnarray}

\newpage

\begin{center}

The SO$_{6} \supset$ SO$_{5}$  isoscalar factors

\vspace{0.5in}

\[
 \left( \begin{array}{cc} \left[\sigma+\frac{1}{2},
 \frac{1}{2}\frac{1}{2}\right] & [100] \\
 \left[\tau+\frac{1}{2}, \frac{1}{2}\right] & [10] \end{array} \right|
 \left. \begin{array}{c} \left[\sigma'+\frac{1}{2}, \frac{1}{2} \frac{1}{2}
 \right] \\
 \left[\tau+\frac{1}{2}, \frac{1}{2}\right] \end{array} \right)
\]

\begin{tabular}{cc}
 $\sigma'=\sigma+1 $ & $\sigma'=\sigma-1$ \\
 $ \left[\displaystyle{\frac{(\sigma-\tau+1)(\sigma+\tau+5)}
 {2(\sigma+1)(\sigma+3)}} \right]^{\frac{1}{2}}$ &
 $ \left[\displaystyle{\frac{(\sigma-\tau)(\sigma+\tau+4)}
 {2(\sigma+2)(\sigma+4)} } \right]^{\frac{1}{2}}$ \\
\end{tabular}

\vspace{0.5in}

\[
\left( \begin{array}{cc} \left[\sigma+\frac{1}{2},
 \frac{1}{2}\frac{1}{2}\right] & [100] \\
 \left[\tau+\frac{1}{2}, \frac{1}{2}\right] & [10] \end{array} \right|
 \left. \begin{array}{c} \left[\sigma'+\frac{1}{2},
 \frac{1}{2}\frac{1}{2}\right] \\
 \left[\tau_1 \tau_2 \right] \end{array} \right)
\]

\begin{tabular}{c|cc}
 $[\tau_1 \tau_2]$ & $\sigma'=\sigma+1$ & $\sigma'=\sigma-1 $ \\ \hline

 $\left[\tau+\frac{3}{2},\frac{1}{2}\right]$ &
 $\left[\displaystyle{\frac{(\sigma+\tau+5)(\sigma+\tau+6)(\tau+1)}
 {2(\sigma+1)(\sigma+3)(2\tau+5)} } \right]^{\frac{1}{2}}$ &
 $-\left[\displaystyle{\frac{(\sigma-\tau)(\sigma-\tau-1)(\tau+1)}
 {2(\sigma+2)(\sigma+4)(2\tau+5)} }\right]^{\frac{1}{2}}$ \\

 $\left[\tau+\frac{1}{2},\frac{3}{2}\right]$ & 0 & 0 \\

 $\left[\tau+\frac{1}{2},\frac{1}{2}\right]$ &

$-\left[\displaystyle{\frac{(\sigma+\tau+5)(\sigma-\tau+1)(\tau+3)}{2(\sigma+1)(\sigma+3)
 (3\tau+2)(2\tau+5)} } \right]^{\frac{1}{2}}$ &
 $ \left[\displaystyle{\frac{(\sigma+t+4)(\sigma-\tau)(\tau+2)}
 {2(\sigma+2)(\sigma+4)(3\tau+2)(2\tau+5)} }\right]^{\frac{1}{2}}$ \\

 $\left[\tau-\frac{3}{2},\frac{1}{2}\right]$ &
 $ -\left[\displaystyle{\frac {(\sigma-\tau+2)(\sigma-\tau+1)(\tau+3)}
 {2(\sigma+1)(\sigma+3)(2\tau+3)} } \right]^{\frac{1}{2}}$ &
 $ \left[\displaystyle{\frac{(\sigma+\tau+4)(\sigma+\tau+3)(\tau+3)}
 {2(\sigma+2)(\sigma+4)(2\tau+3)} }\right]^{\frac{1}{2}}$ \\
\end{tabular}

\end{center}

\newpage

Figure captions

\begin{description}
\item{Fig.\ 1.} Comparison between calculated levels using eq.\ (17) and
 experimental
 energy levels for the even--even $^{120-126}$Xe isotopes.

\item{Fig.\ 2.} Comparison between calculated levels using eq.\ (17) and
 experimental
 energy levels for the even--even $^{128-132}$Xe isotopes.

\item{Fig.\ 3.} Comparison between calculated levels using eq.\ (18) and
 experimental energy levels for the even--odd $^{127-129}$Xe isotopes. The
even--odd nuclei are constructed by coupling the neighboring even--even
core to a valence neutron.

\item{Fig.\ 4.} Comparison between calculated levels using eq.\ (18) and
 experimental energy levels for the even--odd $^{131-133}$Xe isotopes. The
even--odd nuclei are constructed by coupling the neighboring even--even
core to a valence neutron.

\item{Fig.\ 5.} Comparison between calculated levels and
experimental energy levels for the even--odd $^{131-135}$Ba isotopes. For the
 excited [$\tau_{1}\tau_{2}$] states, the band-head states are compared with
experimental ones.

\item{Fig.\ 6.} The systematic shift of the ground band as a function of
 mass number for Xe isotopes.

\item{Fig.\ 7.} The systematic shift of the ground band as a function of
 mass number for Ba isotopes.

\item{Fig.\ 8.} (a) The $SO_{6}$ spectrum calculated
with the parameters are
$g_{6}$=-73.3 (keV), $g_{5}$= 71 (keV) and $g_{I}$=11.9 (keV);
(b) The comparison of the experimental spectrum of $^{126}$Xe
(lower numbers) and the spectrum of
$SO(6)$ plus a perturbative pairing term (upper numbers)
(i.e., eq.\ (26)).
The parameters here are $g_{6}$=-73.3 (keV), $A'$=80 (keV). $B'$=0.77 (keV) and
$g_{I}$=11.9 (keV).

\end{description}

\end{document}